\documentclass[%
amsmath,%
amssymb,%
prb,
showpacs
]{revtex4}

\usepackage{dcolumn}% Align table columns on decimal point
\usepackage{setspace}
\usepackage{url}
\usepackage{graphicx}% Include figure files
\usepackage{bm}% bold math

\newcommand{\eqn}[1]{Eq.\,(\ref{#1})}
\newcommand{\eqns}[1]{Eqs.\,(\ref{#1})}
\newcommand{\noeqn}[1]{(\ref{#1})}
\newcommand{\fig}[1]{Fig.\,\ref{#1}}

\newcommand{\etal}{{\em et al.}}

\newcommand{\hide}[1]{}

\begin{document}

\title{%
  Mathematical Models of Progressive Collapse and the Question of
  How Did the World Trade Centers Perish
}
\author{Charles M. BECK}
\email{beck.charles_m@yahoo.com}
%%%%%%%%%%%%%%%%%%%%%%%%%%%%%%%%%%%%%%%%%%
%%%%%%%%%%%%%%%%%%%%%%%%%%%%%%%%%%%%%%%%%%
%% md5-author-id:                       %%
%%  e48011b3b7250c88c0670ab42f34470b    %%
%%%%%%%%%%%%%%%%%%%%%%%%%%%%%%%%%%%%%%%%%%
%%%%%%%%%%%%%%%%%%%%%%%%%%%%%%%%%%%%%%%%%%
\affiliation{Non-affiliated}
\date{2 October, 2007}

\begin{abstract}
  We derive discrete and continuous class of mathematical models
  that describe a progressive collapse in a fictional one-dimensional structure,
  where we consider plastic and elastic types of collisions.
  We examine static (collapse initiation lines, derived
  from the ultimate yield strength of the structural steel)
  and dynamic (duration of collapse, computed using mathematical models)
  features of events that comprised the collapse in WTC 1 and 2.
  We show that $(a)$, the dynamic and static
  aspects of the collapse are mutually consistent and weakly dependent on
  the class or type of mathematical model used,
  and $(b)$, that the NIST scenario, in which the buildings collapse
  after a sequence of two damaging events (airplane impact and subsequent ambient fires),
  is inconsistent with respect to the structural strength of the buildings.
  Our analysis shows that the force that resisted  the collapse in WTC 1 and 2
  came from a single structural element, the weaker perimeter columns,
  while the second structural element, the stronger core columns, did not contribute.

  We derive continuous model to examine the collapse of WTC 7 and, again,
  find consistency between the static and dynamic features of
  collapse.
  We show that the collapse of WTC 7 to the ground
  is inconsistent with the FEMA/NIST scenario, by which the collapse was due to
  gradually worsening conditions in the building.
  Instead, we find that for the conditions in the building to change, from being
  compromised by heat to being on the verge of collapse, yet another damaging
  event is necessary, the magnitude of which is greater than that of the heat.

  We discuss two non-obvious inconsistencies between the mathematical models
  of progressive collapse based on the NIST scenario, and the practical
  realizations of collapse in WTC 1 and 2:
  $(i)$, the average avalanche pressure is 3 orders of magnitude smaller than
  the pressure the vertical columns are able to withstand, and $(ii)$, the intact
  vertical columns can easily absorb through plastic deformation the energy of the
  falling top section of the WTCs.
  We propose collapse scenario that resolves these inconsistencies, and is in agreement
  with the observations and with the mathematical models.
\end{abstract}

\maketitle

\section{Introduction}

On September 11$^{th}$, 2001, World Trade Centers (WTC) 1 and 2 were each hit by a
commercial airplane.
Approximately 56 minutes after being hit WTC 2 collapsed to the ground in what appears
to be a single-front avalanche, which started in the upper half of the building and reached
the ground floor in $_2T_c \simeq 11$ s.
WTC 1 followed the similar destiny some 50 minutes later, where its collapse lasted
$_1T_c \simeq 13$ s.
In 2002, Federal Emergency Management Agency (FEMA) published a WTC performance
study~\cite{fema:2002} that concluded that the buildings were built properly.
National Institute for Standard and Technology (NIST) followed with
its own report on collapse in 2005~\cite{nist:2005}.
The NIST report agreed with the FEMA findings that the buildings did not have design flaws.
It dealt mainly with the conditions throughout the impacted floors and how the
damage from the impact and subsequent fires could have compromised the load carrying
capabilities of the central core of the building in the impact zone.

In this report we use mathematical models of progressive collapse
with the structural parameters of the buildings to examine the
consistency between static and dynamics features of collapse.
Here, under the static features we assume the conditions in the building
on the verge of collapse, while under the dynamic we assume the
observed duration of collapse as calculated using the mathematical model of
progressive collapse.

The report is organized as follows.
In Sec. II we present the mathematical models
of progressive collapse, where we introduce two classes (discrete and continuous)
and two types (elastic and plastic) of models.
We briefly discuss the underlying physics and how the structural
parameters enter each class of the models.
In Sec. III we discuss the structural parameters of WTC 1 and 2 and introduce
their two main load-bearing structures, the weaker perimeter columns (PCs)
and the stronger core columns (CCs).
In Sec. IV we discuss the NIST scenario of collapse, and show how it is inconsistent
with the static features of collapse (collapse initiation lines) unless
one neglects the contribution of CCs.
In Sec. V we present our conclusions, while in Sec. VI we discuss some less obvious
assumptions.
One group assumptions is hidden behind the one-dimensional formulation of progressive
collapse, while the other behind, so-called, rigidity assumption, by which the range
of elastic/plastic deformation of a column does not spread beyond the stretch of a
single storey.
In Appendix A we present mathematical model of collapse of WTC 7 and analyze it
in the same way as it was done for WTC 1 and 2.

\section{Mathematical Models of Progressive Collapse}

The World Trade Centers 1 and 2 were complex three-dimensional structures, yet it appears
that the progressive collapse in each of them was one-dimensional:
an avalanche formed at the top of the impact zone and propagated to
the ground.
Based on that fact, we describe two classes of the mathematical models
that were proposed for description of the progressive collapse of the buildings:
continuous and discrete.
An assumption, both models critically depend on, is that the avalanche front -
a fictional plane that moves with the velocity of the avalanche and through
which the building underneath joins the avalanche - maintained its
shape (flatness, being parallel to the ground) for the whole duration of collapse.

Less important assumptions are those regarding the loss of mass and momentum
and the rigidity.
As for the losses, in the case of a massive building, such as the WTCs, one could expect
lateral (material leaking on the sides) and longitudinal
(significant variation in resistive force along the avalanche front prevents formation
of a sufficiently compacted layer that would be able to crush strong vertical columns
in its path).
The rigidity assumption, on the other hand, requires that the interaction between
the avalanche and the building is localized only to the point of contact, and not
any further into the building beneath.
In our exposition of the mathematical models we
thus assume no losses of any kind, and a reasonable rigidity in the building,
all of which contributed to the apparent one-dimensionality of the avalanche propagation.

\subsection{Continuous Models}

For the purpose of our models we assume that the avalanche propagates
along the $z$-axis, where the positive direction is along the direction
of Earth's gravity.
The building in which the avalanche propagates thus initially stretches
from 0 (top) to H (bottom), and its longitudinal mass density
is described by $\mu=\mu(Z)$, where $M=\int_0^H\,dZ'\ \mu(Z')$ is the
total mass of the building.

In the continuous class of mathematical models of progressive collapse, the avalanche
at position $Z$ consists of two objects moving together: the top section,
and the avalanche front.
The top section is the part of the building initially stretching from 0 to $Z_0$,
where $Z_0$ is the point at which the avalanche started.
The avalanche front is a zero-size object that contains all the mass the avalanche
has collected during its fall.
The total mass of the avalanche is thus the mass of the top section
and the mass of the avalanche front, and is given by $m(Z) = \int_0^Z\,dZ'\ \mu(Z')$.
The destruction of the building occurs at the interface between the
avalanche front and the stationary part of the building, where all the
transfer of mass and momentum between the two occurs.

We consider the perfectly inelastic (plastic) collisions between the avalanche
front and the building first.
Here, the rate of change of momentum of the avalanche follows from elementary
Newtonian dynamics, and is given by
\begin{equation}
  \label{eq:nc}
  \frac{d}{dt} \left( m(Z) \, \dot Z \right) = m(Z) \, g + {R^*}(Z),
\end{equation}
where $\dot Z = dZ/dt$ is the velocity of the avalanche.
In \eqn{eq:nc} the term ${R^*} = {R^*}(Z)$ comes from the average resistive force
produced by the building.
It measures how does the building oppose its own destruction by the avalanche,
and is intimately related to, so called, the yield strength of the
structural members that provide the vertical support to the building.
The term $m(Z) \, g$ is the usual gravitational force, where $g$ is Earth's gravity.
In this formulation the resulting equation resembles that of a falling
water drop~\cite{ews2001,krane1981} or a falling chain.~\cite{keiffer2001,thornton:2004}

The collisions between the avalanche front and the building can also be
considered as perfectly elastic.
This assumption leads to the Lagrangian formalism, which requires
that all relevant energies be expressible in terms of a small set of
configuration coordinates and their time derivatives (generalized velocities).
For this particular problem position of the avalanche, $Z$, and
its velocity, $\dot Z$, are the most obvious choice.
The relevant energies necessary for Lagrangian are as follows:
$K(Z,\dot Z) = \frac 1 2 \, m(Z) \, \dot Z^2$ is the kinetic energy, while
$U(Z) = - m(Z) \, g \,Z - \int_Z^H dX \, \mu(X) \, g \, X$ is its
potential energy.
Additionally, we need energy associated with the structure of the building,
$L = - \int^Z\,dX \ R^*(X)$, as a result of which the resistive
force appears.
In Lagrangian formulation the equation of motion follows from
$\frac {d}{dt} \, \partial{\cal L}/\partial \dot Z = \partial{\cal L}/\partial Z$,
where ${\cal L} = K - U - L$ is the Lagrangian:
\begin{equation}
  \label{eq:c}
  \frac{d}{dt} \left( m(Z) \, \dot Z \right) = m(Z) \, g
  + \frac 1 2 \, \mu(Z) \, \dot Z^2 + R^*(Z).
\end{equation}
A practical physical realization of such system is a standard
textbook problem~\cite{wy2005,thornton:2004} of
a falling chain.
Please note, \eqn{eq:c} can also be derived from \eqn{eq:nc} by considering
not only a mass transfer but also a transfer of momentum.~\cite{thorpe1962,tiersten1969}

\eqns{eq:nc} and \noeqn{eq:c} can be jointly written as
\begin{equation}
  \label{eq:cnc}
  \frac{d}{dt} \left( m(Z) \, \dot Z \right) =
  m(Z) \, g  + \frac 1 2 \, \epsilon \, \mu(Z) \, \dot Z^2 + R^*(Z),
\end{equation}
where we introduce an additional parameter, elasticity, $\epsilon\in[0,1]$,
which describes the type of collisions between the building and the avalanche front,
of which plastic and elastic collisions are just the extremae,
$\epsilon=0$ and $\epsilon=1$, respectively.
One can envision obtaining the value of $\epsilon$ by fitting the
actual progression of the collapse in time to that predicted by the
model~\eqn{eq:cnc}.

To simplify the analysis of progressive collapse further we assume that
the building is homogeneous height-wise, i.e., $\mu = M / H$.
Finally, we convert \eqn{eq:cnc} to dimensionless form where we use
as the unit of length the height of the building $H$, and as the
unit of time the free fall time from height $H$, $T=\sqrt{2\, H / g}$.
The position of the avalanche $z=Z/H$ in time $\tau=t/T$ is described as a solution
of the following ordinary differential equation:
\begin{equation}
  \label{eq:cnc:ode}
  z''(\tau) = 2 - \left(1 - \frac{\epsilon}{2}\right) \,
  \frac{z'^2(\tau)}{z(\tau)} + \frac{2}{z(\tau)}\,\frac{R^*(z)}{M\,g},
\end{equation}
where prime denotes the differentiation with respect to dimensionless time $\tau$.
We elaborate on the initial conditions $z(0)$ and $z'(0)$ later, while we
discuss the resistive force $R(z)$ next in the context of the discrete models.

\subsection{Discrete Models}

In the discrete models the building is represented by a periodic structure
of stacked stories,~\cite{bazant2006}
where each storey consists of a floor and a number of vertical
columns which keep the stories at their positions
$Z_F$, where $F=1,..F_T$ is the floor number, while $F_T$ is the total number
of floors in the building.
The height of each storey is $\Delta H = H/F_T$.

Just like in one-dimensional continuous model \eqn{eq:cnc}, the avalanche
in discrete model propagates only in vertical direction.
Let us assume that the consuming of a storey by the avalanche comprises of
three phases, {\em resistive}, {\em free fall} and
{\em adsorption},
where each phase takes a definite amount
of fractional floor height, $\lambda_1$ for resistive,
$\lambda_2$ for free fall, and $\lambda_{\infty}$ for adsorption.
Obviously, $\lambda_1 + \lambda_2 + \lambda_\infty = 1$.
The fractional height $\lambda_\infty$ is sometimes called the compaction
ratio.~\cite{bazant}
During first two phases the avalanche moves without acquiring the mass,
where the resistance of the vertical columns is $R^1$ and $R^2\equiv0$, respectively.
After dropping by the fractional height $(\lambda_1+\lambda_2)$ the avalanche
makes a discontinuous jump by $\lambda_\infty$ and adsorbs
the compacted storey in the process.
The progressive collapse of the $j$-th storey of
mass $\Delta m = M/F_T$ is described by the following
set of dimensionless equations:
\begin{subequations}
  \label{eq:d}
  \begin{equation}
    \label{eq:d:1}
    \left.
      \begin{array}{lcl}
        a_k &=& 2 + \frac{2}{j-1} \, \frac{R^k_{j-1}}{M\,g}\\
        u_k^2 &=& u_{k-1}^2 + 2 \, \frac{\lambda_k}{F_T} \, a_k\\
        \Delta \tau^k   &=& \frac{u_k - u_{k-1}}{a_k}
      \end{array}
    \right\} \mbox{ for } k=1,2,
  \end{equation}
  \begin{equation}
    \label{eq:d:2}
    v_j = \frac {j-1 + \frac{\epsilon}{2}}{j} \, u_2,
  \end{equation}
\end{subequations}
where $j = F_T - F$, with $F$ being the floor number of the current avalanche position.
Here, $u_0 = v_{j-1}$ is the velocity of the avalanche at the beginning of the
storey, while $\tau_j -\tau_{j-1} = \Delta \tau^1  + \Delta \tau^2$ is how long it took
the avalanche to transverse the storey.
In \eqn{eq:d:2} we introduce a parameter $\epsilon$ in the same
way it appears in the continuous model \noeqn{eq:cnc},
\begin{equation}
  \label{eq:d:e}
  \epsilon = 2 \, \frac{u'}{u_2},
\end{equation}
where $u'$ is the average velocity of the compacted storey at the moment it
is adsorbed by the avalanche, the velocity of which is $u_2$, see \eqn{eq:d:1}.

In \eqns{eq:d:2} and \noeqn{eq:d:e} one can see that the rigidity assumption
put forth by Bazant \etal\cite{bazant2000} is expected to hold, at best, only for
the height scales greater than the storey height, but not for smaller.
The compaction of the storey by the avalanche begins the moment the avalanche enters
the storey, and by the time the avalanche reaches the position $\lambda_1+\lambda_2$
some of the storey's mass is already moving with the velocity of
the avalanche $u_2$, so in fact $\epsilon>0$.

\subsection{Resistive Force: discrete vs. continuous}

The resistive force $R$ is the force the building opposes its own destruction
by the avalanche.
Bazant \etal\cite{bazant2006} call it the ``crushing force,'' and for the most part of their
calculation they assume that its magnitude is a constant with respect to height.
For our investigation we relax that assumption by allowing
$R$ to vary with height in a simple linear fashion,
\begin{equation}
  \label{eq:R}
  f(z) = -\frac{R(z)}{M\,g} = r + s \cdot z,
\end{equation}
where $M$ is the total mass of the building, and $z=Z/H$ is the scaled height.
As we will see later, when we discuss the load-bearing capacity of the structure
of WTC 1 and 2, this will turn out to be more accurate description.

Let us now discuss the resistive force within the discrete (floor) model.
We posit that the building is rigid, i.e., that the floor that avalanche
is currently penetrating sits on an infinitely strong structure.
With that assumption the interaction between the building and the avalanche
is localized to the floor of contact between the two and not beyond.
Now, as the avalanche enters the storey it first encounters the resistive force from
the vertical columns, call it $R^1$.
Over the fractional length $\lambda_1$ the vertical columns
maintain ultimate yield force under compression.
In other words, for the first phase of floor destruction
the fractional distance $\lambda_1$ corresponds to the yield strain $\lambda_Y$,
while $R^1$ corresponds to the ultimate yield force.

Assuming that avalanche has passed the fractional distance $\lambda_1$,
the columns fail.
The failure mode of the column is debatable:
if the columns  are compressed by the avalanche than they fail by buckling. However, if
the avalanche front consists of crushed material, so that the ends of the vertical
columns facing the avalanche are free, then the failure mode may be bending, as well.
Either way, we assume that during this phase the column does not offer any resistance.
In other words, for the fractional distance of $\lambda_2$ the avalanche falls freely,
$R^2 \equiv 0$.

The last phase of propagation occurs when the avalanche reaches the fractional distance
of $\lambda_1+\lambda_2 = 1 - \lambda_\infty$, that is, it compacts the storey.
We assume that then the position of the avalanche makes a discontinuous jump
from $1 - \lambda_\infty$ to 1 (bottom of the current storey) and adsorbs the mass
of the storey as discussed earlier, see \eqn{eq:d:2}.

The resistive force $R^*=R(r^*,s^*)$ that enters the continuous models
is related to $R=R(r,s)$ in the discrete model via the energy argument.
By that argument the energy $\Delta L$ over the fractional length
$\lambda_1+\lambda_2$ is the same in both models, yielding
\begin{subequations}
  \label{eq:rc:all}
  \begin{equation}
    \label{eq:rc:1}
    r^* = \frac{\lambda_1}{1-\lambda_\infty} \, r,
  \end{equation}
  \begin{equation}
    \label{eq:rc:2}
    s^* \simeq \frac{\lambda_1}{1-\lambda_\infty} \, s.
  \end{equation}
\end{subequations}
As we have two descriptions of the resistive force, we choose the discrete
as the primary.
Thus, whenever we talk about the parameters of the building we always assume
those entering the discrete model, thus $(r,s)$.
If we use continuous models for what ever reason then we imply that
the nominal, or discrete, values $(r,s)$ were converted to $(r^*,s^*)$
via~\eqn{eq:rc:all} before commencing the computations.

A question may arise if the two models for the same nominal values of $r$ and $s$
give the same or, at least, similar results.
The question is not so much relevant from the computational point of view:
recursive relations~\noeqn{eq:d} comprising the discrete model are actually simpler
to numerically implement than the ODE comprising the continuous model~\noeqn{eq:cnc:ode}.
On the other hand, the continuous model allows us to use a number of theoretical
tools (Lagrangian formalism, energy analysis, \& c.).
The main difference between the models is in the distance the avalanche
propagates.
In the continuous model the avalanche covers the full height of each storey, $\Delta H$.
In the discrete model, on the other hand, in each storey a section of
fractional length $\lambda_\infty$ is excluded from propagation,
so the effective height per story is $(1-\lambda_\infty) \, \Delta H$.
For the quantity we are interested in, which is the duration of collapse
or {\em collapse time}, we expect both models to be similar for the following
reason.
The collapse time $t$ depends on the velocity $u$ and the distance $d$ as
$t \sim d /u$.
When we change from discrete to continuous model the path increases but so does the velocity
as there is more energy available.
In the lowest order of approximation the two effects are thus expected to
counteract each other.
In fact, we use the spread between the two models as a measure of uncertainty
when determining $r$ and $s$ that yield desired (observed) collapse time.

On the other hand, by comparing plastic ($\epsilon\equiv0$) to
elastic case ($\epsilon\equiv1$) the following is obvious: given $(r,s)$
the collapse time is shorter in elastic than in plastic case.
This difference, as is common knowledge, comes from the energy dissipation inherent
in plastic model.
In spirit of ``no-energy-and-momentum-left-behind,'' by which we try to neglect as many
losses as possible to obtain the lower estimates on collapse times,
we assume $\epsilon\equiv1$ for most of the computations that follow.

\subsection{Bazant's model}

Bazant \etal\cite{bazant2006} proposed the following mathematical
model to describe the progressive collapse in World Trade Center 2,
\begin{equation}
  \label{eq:bazant:1}
  \left(1 - \lambda_\infty \right) \,
  \frac{d}{dt} \left( m(Z) \, \dot Z \right) = m(Z) \, g + {R}(Z),
\end{equation}
where, as before, $Z$ is the position of the avalanche front and $R(Z)$ is the
average floor ``crushing'' force.
Additionally, the resistive force that enters the computations is
to be halved along the whole length of the building due to the ``heat'' effects
(see p.15 on-line edition, top paragraph).
Finally, for the most part of their presentation the authors assume that the
computational value for the parameters $(r,s)$ of the resistive force
are $s=0$ and $r=0.15$
(for crushing energy of $W_f = 2.4$ GNm, $\Delta H=3.7$ m, where the total mass
of the building is $M=450\cdot10^6$ kg).
Based on their modeling the authors concluded ({\em sic!}) that there was nothing
unusual in the way the building collapsed.

The rest of this paper is to show that the alternative interpretation
is possible.
As already discussed, the scaling of kinetic term on the right-hand-side
in \eqn{eq:bazant:1} is unnecessary to equate the discrete
to the continuous model.
In fact, this procedure is equivalent to boosting the force that accelerate
the avalanche by a factor of $(1 - \lambda_\infty)^{-1}$.
Similarly, the model \eqn{eq:bazant:1} is non-conservative to embody rigidity
assumption against which, the reader might recall, we argued.
Finally, no attempts were made on authors' behalf to correlate the information about
the status of the building to the values of the parameters that enter the model.
Heat argument that the authors used to halve the magnitude of the resistive force in
the whole building is in direct contradiction with the NIST report, which explicitely
states that no temperatures higher than 200$^o$C were measured anywhere in the building
below the impact zone.

Overall, it appears that the authors tried to compensate for slowness of their
non-conservative model by boosting the energy $U+L$ and arbitrary halving
the strength of the building.
For these reasons, we exclude Bazant's model from the analyses that take the
rest of the paper.

\section{Structural parameters of the World Trade Centers}

World Trade Centers 1 and 2 were designed as the external tubular frame around
the strong central core.~\cite{web:wtc}
Tubular frame consisted of 236 perimeter columns (PC), while
the central core consisted of 51 core columns (CC), all from structural
steel of nominal strength 36-100 KSI.
These provided load-bearing support to the whole structure, while the
combination of steel trusses and concrete panels stretching between the
two groups of columns provided actual floors for occupants to dwell on.
We recall that the destruction is determined by the ultimate, rather than the
nominal strength of the vertical columns.

PCs, the external dimensions of which were 14''-by-14'', were made of structural
steel the yield strength of which varied from nominal 36 KSI
(ultimate 58 KSI) and thickness $1\over4$'' at
the top of the building to nominal 100 KSI (ultimate 110 KSI)
at the bottom.\cite{nist:2005}
The thickness of the steel plates at the bottom, to the best of author's
knowledge, is not yet publicly available.
CCs were made of structural steel which varied from nominal 36 KSI (ultimate 58 KSI)
at the top to nominal 42 KSI (ultimate 60 KSI)  at the bottom.
Neither the dimensions or the thickness of the steel plates used for CCs,
to the best of the author's knowledge, are yet publicly available.
The NIST report~\cite{nist:2005} claims that there were two types of
CCs: ``standard'' and ``massive,'' where the four ``massive'' columns,
one at each core's corner, ``together provided 20\% of load-bearing
capacity of the core.''
This was recently proven to be misleading:\cite{911:core}
information released to the public shows that of 51 CCs at least 16 (along two longer
sides) were of external dimensions 22''-by-55'' and of thicknesses
up to 5''.
To obtain the values for our models we proceed as follows.
For each structural element, the PCs and the CCs, we find the scaled ultimate
yield force $f=f(z)$, at the top ($z=0$) and at the bottom ($z=1$).
The parameters $r$ and $s$ are then obtained using
\begin{subequations}
  \label{eq:rs:def}
  \begin{equation}
    \label{eq:rs:def:r}
    r = f(0),
  \end{equation}
  \begin{equation}
    \label{eq:rs:def:s}
    s = f(1)-f(0).
  \end{equation}
\end{subequations}

For the PCs we assume that the thickness of the plates at the bottom was $1\over2$'',
yielding
\begin{equation}
  \label{eq:pc:yield}
  f_{PC}(z) = \frac{R_{PC}}{M\,g} \simeq 0.2 + 1.2 \cdot z,
\end{equation}
where $M = 4.5\cdot10^8$ kg is the estimated mass of the building, while
$g$ is gravity.

As for the CCs, we do base our expectations
on the following: all 51 column were of dimensions 22''-by-55'', while
their thickness and strength varied from 1$1\over4$'' and nominal 36 KSI
(ultimate 58 KSI) at the top, to 5'' and nominal 42 KSI (ultimate 60 KSI)
at the bottom.
This yields,
\begin{equation}
  \label{eq:cc:yield}
  f_{CC}(z) = \frac{R_{CC}}{M\,g} \simeq 0.6 + 1.5 \cdot z.
\end{equation}
Please note, the assumption of all 51 columns being the same appears to
be corroborated by the floor plans:~\cite{nist:2005} all CCs appear to
have the same footprint.
Also, while this estimate is purely hypothetical, it turns out that,
it is also irrelevant for the analysis of collapse that follows.

We observe that from \eqns{eq:pc:yield} and \noeqn{eq:cc:yield}
the ultimate safety factor of the WTCs is $f_{CC}(1)+f_{PC}(1) \sim 3.5$.
This is a reasonable estimate considering the dimensions of the buildings
and the other safety requirements that entered the structural calculation
(ability to withstand hurricane winds and an airplane impact).

From the properties of structural steel\cite{materials} it is known that the yield
strain under tension and compression are fairly similar, and is $\sim 21-25\%$.
In our model this is represented by $\lambda_1$, which we take to be $\lambda_1 = 0.2$.
The value of compaction limit we take from Bazant,\cite{bazant2006}
$\lambda_\infty=0.2$, which leaves $\lambda_2 = 1-\lambda_1-\lambda_\infty = 0.6$.
From there, $(r^*,s^*)$ in the continuous model are related to
 $(r,s)$ in the discrete values as,
\begin{equation}
  \label{eq:rs:wtc}
  \begin{array}{lcl}
    r^* &=& 0.25 \cdot r,\\
    s^* &=& 0.25 \cdot s.
  \end{array}
\end{equation}

\section{Failure of the NIST scenario}

As is known, WTCs 1 and 2 were subjected to two damaging events, as a result of
which the buildings eventually collapsed in ${_1T}_c=13$ s (WTC 1),
and ${_2T}_c=11$ s (WTC 2).
The first event was an airplane collision in which some of the vertical columns
were cut or otherwise compromised.
We label that event with a letter $M$, for {\em Mechanical}.
The second event was a gradually developing heat damage from ambient and
jet-fuel fires.
We label that event with a letter $H$, for {\em Heat}.
Both types of damage were delivered to each building only in the
primary or impact zone, while the secondary zone (below the primary)
was left intact.
We note that the primary zone in WTC 1 spread over floors $F_0=99$ to $F_1=93$,
while in WTC 2 it spread between the floors $F_0=85$ and $F_1=77$.
Both buildings had the
same number of floors $F_T=110$.
\begin{figure}[h]
  \singlespacing
  \centering
  \includegraphics[scale=0.4,clip]{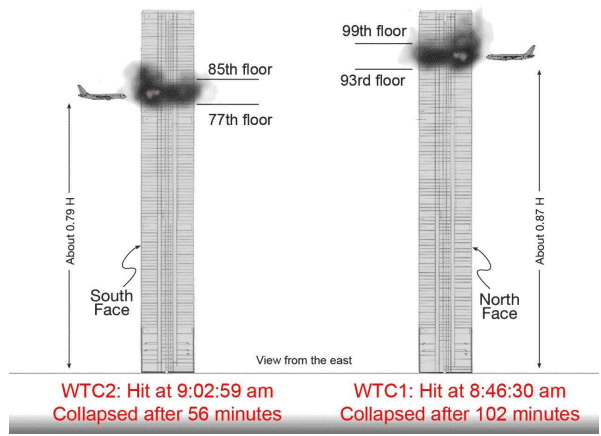}
  \caption{%
    \label{fig:wtc}
    Schematic of airplane collisions with WTC 1 and 2 on September 11,
    2001~\cite{web:wtc:scheme}.
  }
\end{figure}

Let us now formulate, what we call, the NIST scenario for the collapse of
WTC 1 and 2.
For that we need to define first {\em the primary zone damaging event} or NIST($\mu$):
\begin{quote}
  {\bf Definition:} \em A primary zone damaging event, NIST($\mu$), is an incident
  as a result of which the nominal strength of the resistive force in the primary zone
  of the building, $\tilde R$, is reduced by a factor $\mu\ge1$ with respect to
  the value in intact building, $R=R(z)$.
\end{quote}
This is written as,
\begin{equation}
  \label{eq:nist:mu}
  \tilde R(z) = \left\{
  \begin{array}{ll}
    \frac{1}{\mu}\, R(z), & \mbox{ for } z_0 < z < z_1,\\
    R(z), & \mbox{ for } z_1 > z,
  \end{array}
  \right.
\end{equation}
where $z_i = 1 - F_i/F_T$, for $i=0,1$.
The NIST scenario regarding the collapse of WTC 1 and 2 is thus formally expressed
as,
\begin{quote}
  {\bf Definition:} \em
  A NIST scenario of collapse, or NIST($\mu \cdot \nu$), is a sequence
  of two primary zone damaging events to which a building is exposed:
  $M$, with $\mu\ge1$, and $H$, with $\nu\ge1$.
  As a result of the first event, NIST($\mu$), the building does not collapse.
  It collapses only after the second primary zone damaging event,
  NIST($\mu \cdot \nu$), where the duration
  of collapse is known to be $T_c$.
\end{quote}

Let us estimate factors $\mu$ and $\nu$, as they are essential for the NIST scenario.
The NIST report~\cite{nist:2005} documented the damage to the primary zone
fairly well.
WTC 1 suffered frontal hit, and after the aircraft penetrated the building it
exploded inside.
We assume that for WTC 1 we have
\begin{equation}
  \label{eq:mu:wtc1}
  \frac{1}{_1\mu} = 0.5,
\end{equation}
that is, we assume that the aircraft destroyed 50\% of the PCs and of the CCs.
WTC 2, on the other hand, was hit in the corner.
Parts of the airplane flew through the building and most of the fuel
exploded outside the building.
We thus assume that for WTC 2 we have
\begin{equation}
  \label{eq:mu:wtc2}
  \frac{1}{_2\mu} = 0.75,
\end{equation}
that is, we assume that the aircraft destroyed 25\% of the PCs and the CCs.
As for the heat damage, it is reported\cite{nist:2005} that the temperatures of up
to 600$^o$C were measured at some of the locations in the primary zone.
As is known, at that temperature the structural steel loses approximately
one-half of its strength.
We thus have
\begin{equation}
  \label{eq:nu:wtcs}
  \frac{1}{\nu} = 0.5,
\end{equation}
for both buildings.

Now that we have $\mu$ and $\nu$ for each building, we recall that the
collapse initiation occurred at the top of the primary zone,
at position $z_0 = 1 - F_0/F_T$, where for
WTC 1, ${_1z}_0 =0.1$, and for WTC 2, ${_2z}_0 = 0.23$.
This allows us to construct the collapse initiation lines.
The collapse starts at point $z_0$ because the yield force of
the compromised building, $\tilde  R(z_0)$, is not sufficient to resist the
weight of the building above,
\begin{equation}
  \label{eq:damage}
  -\tilde R(z_0) < m(z_0) \cdot g.
\end{equation}
Please observe, the collapse initiation lines are derived from static properties of the
buildings, thus they do not depend on the mathematical model
used to describe the dynamics of collapse.

The collapse initiation lines according to the NIST scenario of collapse
are as follows:
\begin{itemize}
\item
  {\em 0-damage line}:
  \begin{equation}
    \label{eq:0-damage}
    \frac r {{_j z}_0} + s = 1.
  \end{equation}
\item
  {\em M-damage line}:
  \begin{equation}
    \label{eq:M-damage}
    \frac r {_j\mu \cdot {_jz}_0} + \frac{s}{_j\mu} = 1.
  \end{equation}
\item
  {\em M+H-damage line}:
  \begin{equation}
    \label{eq:M+H-damage}
    \frac r {{_j\mu} \cdot \nu \cdot {_jz}_0} + \frac{s}{{_j\mu} \cdot \nu} = 1.
  \end{equation}
\end{itemize}
where $j=1,2$, for WTC 1 and 2, respectively.
Please note, in \eqns{eq:0-damage}-\noeqn{eq:M+H-damage}, $r$ and $s$ represent
the intact building parameters.
Obviously, these lines limit $r$ and $s$ the buildings could have had.
E.g., the buildings do not collapse originally, thus their $r$ and $s$ are
somewhere above the 0-damage line,  \eqn{eq:0-damage}.
Similarly, the building does not collapse after an $M$-event,
thus $r$ and $s$ are somewhere above the $M$-damage line,
\eqn{eq:M-damage}.
On the other hand, the buildings do collapse after a combined $M+H$-event,
thus $r$ and $s$ are somewhere between the $M$-damage line and
the $M+H$ damage line, \eqn{eq:M+H-damage}.

\begin{figure}[htp]
  \singlespacing
  \centering
  \includegraphics[scale=0.7,clip]{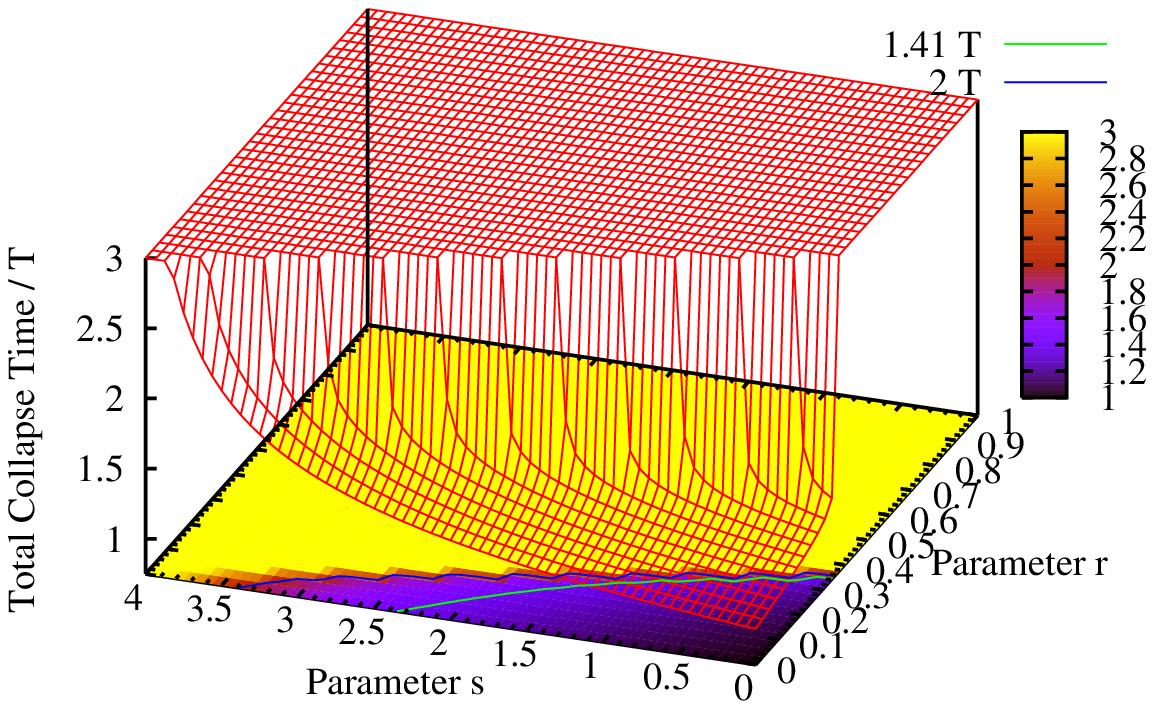}
  \includegraphics[scale=0.7,clip]{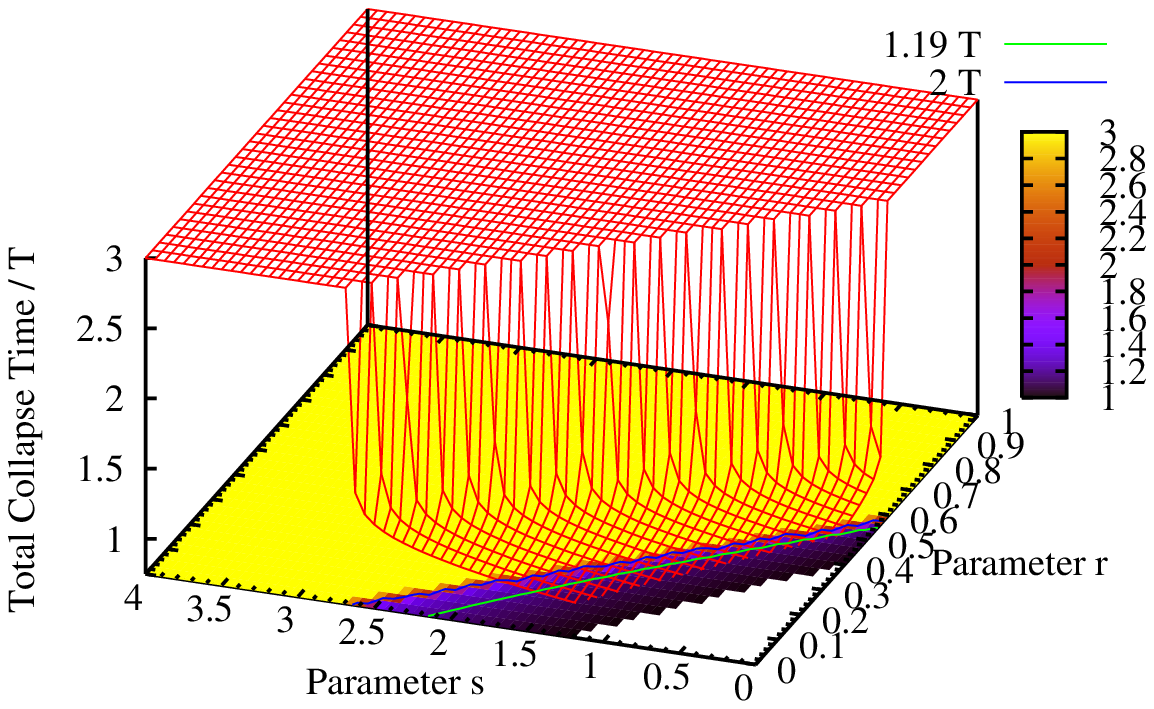}
  \includegraphics[scale=0.7,clip]{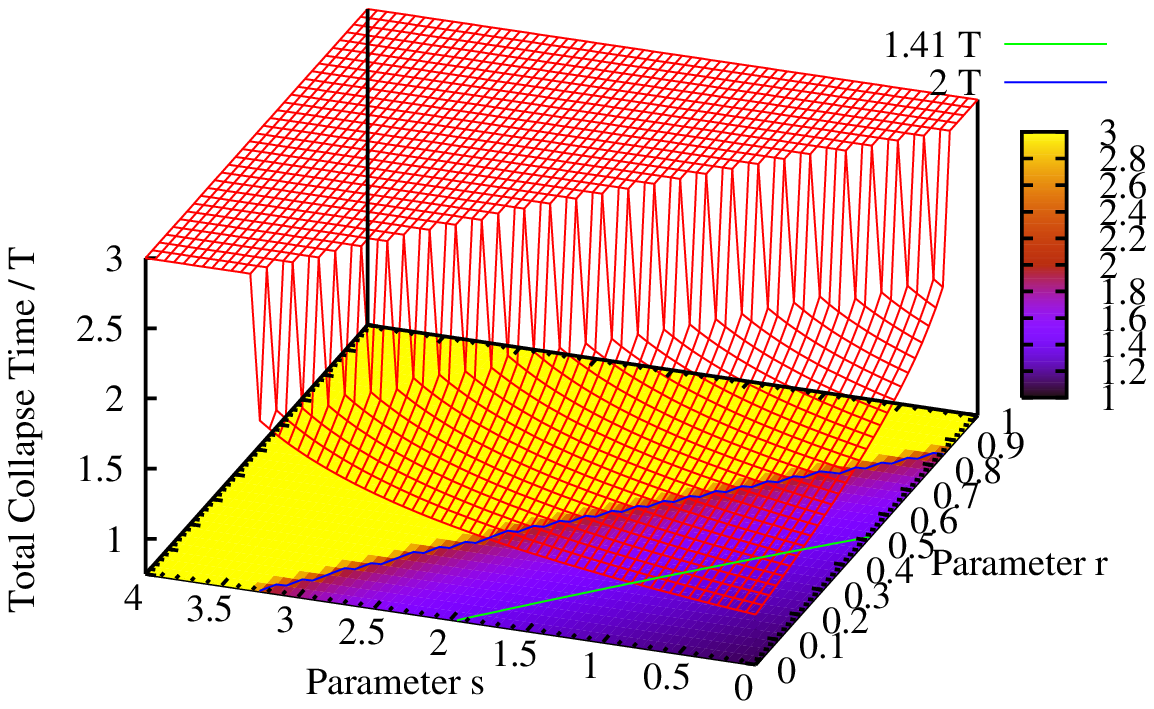}
  \includegraphics[scale=0.7,clip]{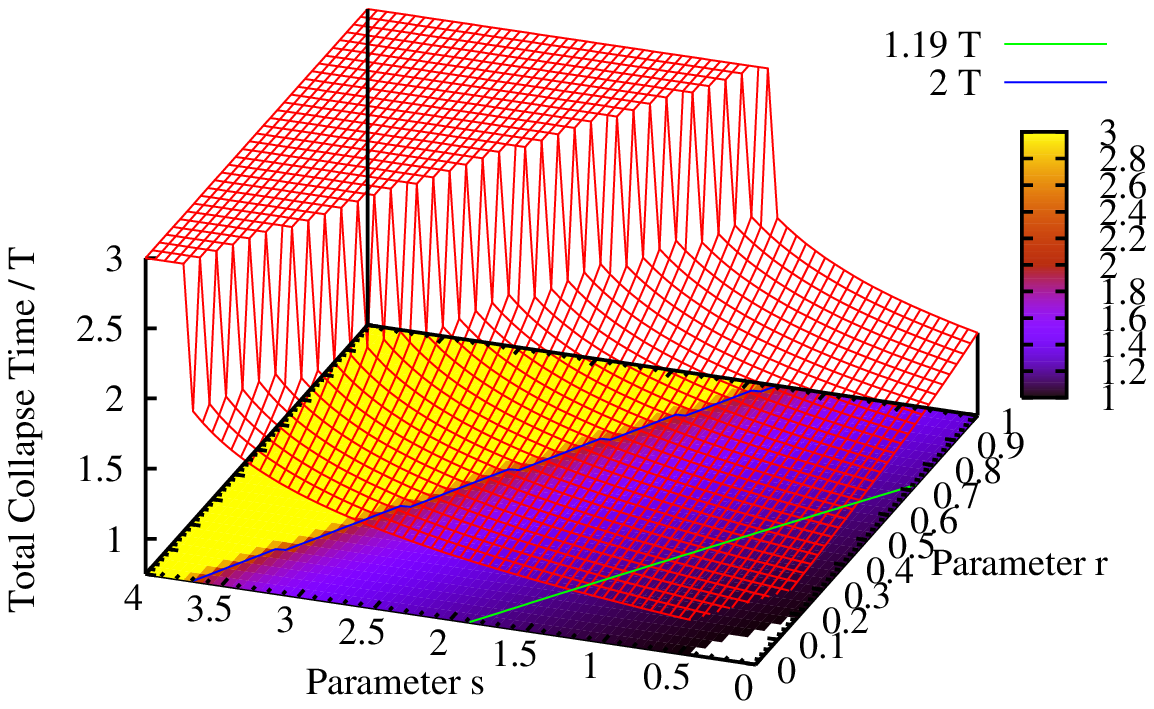}
  \caption{%
    \label{fig:2}
    Collapse time in the $NIST(\mu\cdot\nu)$ scenario,
    in the discrete model (top row) and in the continuous model (bottom row),
    in WTC 1 (left column) and WTC 2 (right column).
    The base contains the contours $T(r,s) = {_j\tau}_c, 2$ for each building.
  }
\end{figure}
\begin{figure}[htp]
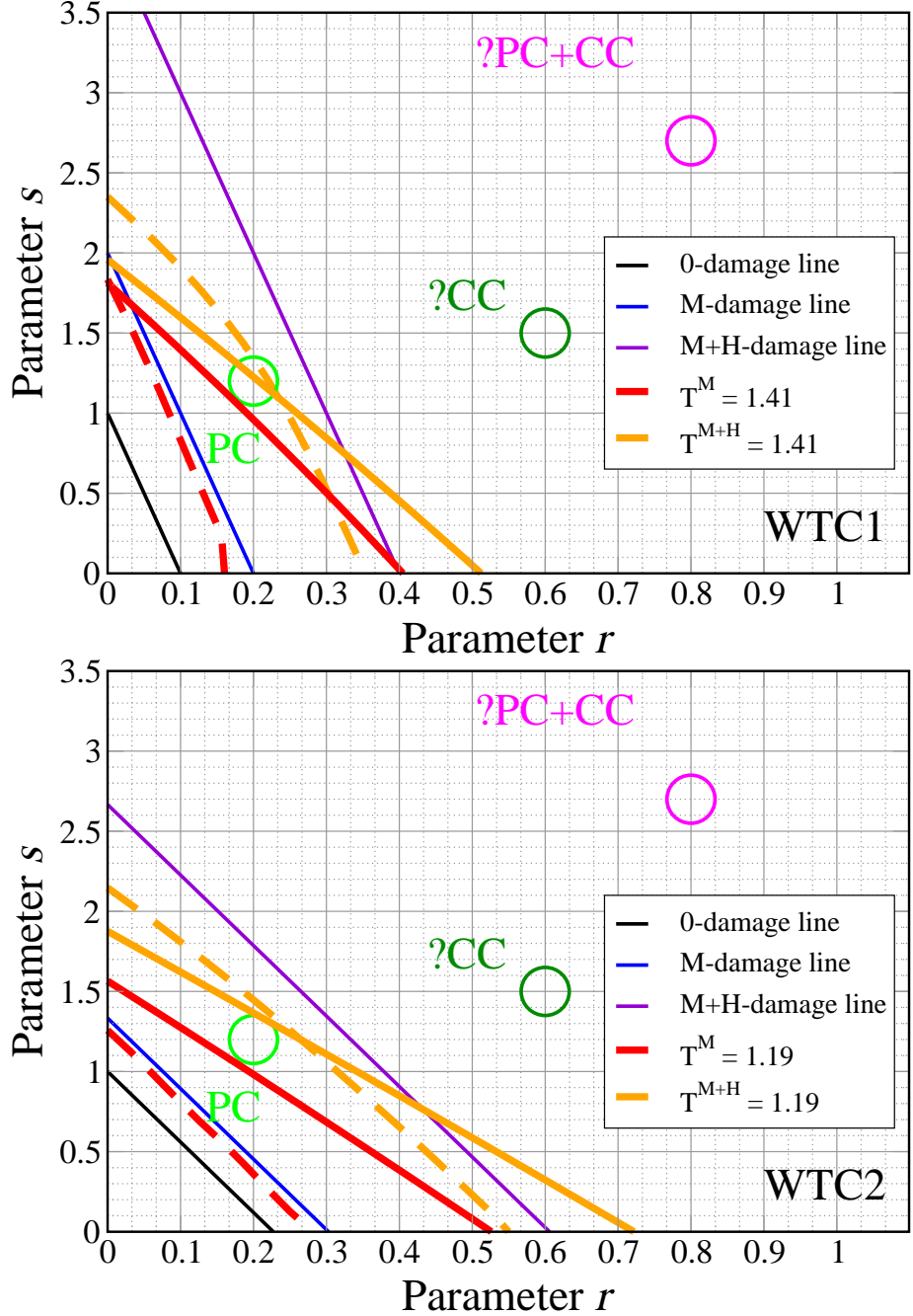

  \singlespacing
  \centering
  \includegraphics[scale=0.5,clip]{fig3a.eps}
  \includegraphics[scale=0.5,clip]{fig3b.eps}
  \caption{%
    \label{fig:3}
    Collapse initiation lines (black from \eqn{eq:0-damage},
    blue from \eqn{eq:M-damage}, purple from \eqn{eq:M+H-damage}),
    with the collapse time $\tau(r,s)=\tau_c$ as calculated
    using the continuous (solid red and orange) and the
    discrete model (dashed red and orange), for $NIST(\mu)$ (red) and
    $NIST(\mu\cdot\nu)$ scenario (orange).
    Both collapse duration lines, $T^{M}=\tau_c$ (red), and
    $T^{M+H}=\tau_c$ (orange), are consistent with the collapse
    initiation lines, $M$-damage line (blue) and $M+H$-damage line (violet),
    respectively.
    The collapse duration lines do not prefer either scenario, and do not
    reveal new information except that the heat damage was below the
    assumed (maximal) heat damage.
    The points representing the structural parameters of the building,
    known ultimate yield force of PCs (green), guessed for CCs (green, with
    question mark) and guessed for their sum (pink, with question mark),
    show that the collapse is consistent with the resistive force coming
    from the PCs only.
  }
\end{figure}
Now we use the mathematical models of progressive collapse and calculate the
collapse duration lines
${_j\tau}^M(r,s) = {_j\tau}_c$ and ${_j\tau}^{M+H}(r,s) = {_j\tau}_c$, for $j=1,2$,
where we assume that the avalanche started at the top of the primary zone, $z(0) = {_jz}_0$,
with zero velocity, $z'(0) = 0$.
The total collapse times are scaled with the free fall time $T = \sqrt{2\,H/g} = 9.22$ s,
from the height of the building, $H=417$ m,
so that ${_2\tau}_c = 1.41$ (${_2T}_c \simeq 13$ s, for WTC 2) and ${_1\tau}_c = 1.19$
(${_1T}_c \simeq 11$ s, for WTC 1).
In our mathematical models we use $\epsilon\equiv1$, as this provides the shorter
collapse times than $\epsilon\equiv0$.
We show in \fig{fig:2} collapse time as a function of building
parameters $r$ and $s$ in each building for the $NIST(\mu\cdot\nu)$ scenario.

In \fig{fig:3} we plot the collapse duration lines (from discrete and continuous model)
for $NIST(\mu)$ and $NIST(\mu\cdot\nu)$ scenarios, the collapse initiation lines
and the points that correspond to the known ultimate
yield strength of the perimeter columns (PCs), the guessed ultimate yield
strength of the core columns (CCs, with question mark), and the
guessed ultimate yield strength of both (CC+PC, with question mark).
We observe that the position of the of the collapse duration lines,
${_j\tau}_c = \tau(r,s)$, varies somewhat with respect to the class of the model
(discrete or continuous), but the difference is not critical.
In fact, we use this spread to estimate the approximate range
of $(r,s)$ that yield the observed collapse time $\tau_c$.

For the $M$-event (mechanical damage from the airplane impact),
we observe that the respective collapse initiation line overlap with the
collapse duration lines.
On its own, the overlap implies that the spontaneous collapse of the building
as a result of mechanical damage only is not impossible.
However, because $r$ and $s$ corresponding to the $M$-event are well below the
estimated strength of the PCs
The PCs in each building's secondary zone appear to have
been non-compromised prior to collapse, so this must be a minimum resistive
force the building could have provided.
Thus, we conclude that the buildings survived the $M$-event, which in fact they did.

For the $M+H$ event (heat damage that followed mechanical damage), in the expected
range of parameters ($r$ and $s$ above that of PCs) the collapse initiation line
does not overlap with the collapse duration lines.
This can be interpreted in two ways:
first one is that the collapse was not spontaneous, while
the second is that the reduction of strength due to heat damage
was smaller than presumed, that is, $\nu < 2$.
We observe that, if the latter were the case, by decreasing $\nu \rightarrow 1$,
the lines for the $M+H$ event (collapse initiation and duration) move toward the same
lines for the $M$ event.
However, the collapse duration lines move ``slower''
than its $M+H$-damage line, so for some $\nu < 2$ the two can overlap.

More importantly, we find that the $r$ and $s$ of the PCs are located
in the region bounded by the collapse duration lines.
This means that in each building the collapse initiation and duration are consistent
with the NIST($\mu\cdot\nu$) scenario being applied to the PCs only,
while the stronger core columns (CCs) are not present at all.
This in turn implies that the NIST scenario is incomplete:
the collapse of the buildings to the ground requires yet another damaging event,
the sole purpose of which is a destruction of the CCs in the secondary zone.
We label this damaging event the ``wave of massive destruction'' (WMD), because of
its catastrophic nature.
Interestingly, the avalanche we have discussed so far can only appear in its wake,
and is thus a result of the WMD rather then the other way around.

\section{Conclusion}

We have determined the static and the dynamic features of a progressive collapse
in the WTCs using the structural properties of the building and the
mathematical models of the avalanche propagation.
We have formally expressed the destruction scenarios proposed by NIST
as a sequence of damaging events in the primary (or impact) zone of each building,
which leave the secondary zone (below) intact.
We have shown that the static and dynamic features of collapse are mutually consistent.
On the other hand, we have demonstrated that the NIST scenarios are inconsistent
with the structural parameters of the building.
More precisely, the features of the avalanche propagation (initiation and duration)
indicate that in their final moments the buildings did not have the core columns (CCs).
We conclude that the buildings did not perish because of combined
mechanical and heat damage to their primary zones,
but because of yet another catastrophic event:
a wave of massive destruction (WMD) that destroyed the CCs, following which
the buildings collapsed to the ground.

\section{Discussion}

A physical situation where a building is being pushed to the ground by the impact of its top
section, where its structural strength comes from a few vertical elements,
is implausible.
Consider the following:
\begin{itemize}
\item[1.] {\bf Destruction of vertical columns:}
Floor plans of the WTCs show floors shaped as flat rectangular doughnuts.
The perimeter columns (PCs) are on the outside, while the core columns (CCs)
are on the inside.
Each floor consists of a concrete surface supported by trusses stretching between
the vertical columns.
If we thus think of the avalanche front as being made of the floor material,
we see that the avalanche front does not stretch horizontally far enough
to reach the PCs and CCs.
In the proposed model of collapse, however, it is implied that $(i)$ the
avalanche front is wide enough to reach both columns, and $(ii)$ provides
sufficient pressure at the edges so that the vertical columns fail.
These assumptions are well hidden in the one-dimensional formulation of our
mathematical models:
all the pressures are integrated over the perpendicular cross section
of the building, and only then the equations of motion are derived.

Consider the average pressure created by the avalanche at the bottom of
the primary zone, $z_1 = 1-F_1/F_T$, which is given by
\begin{equation}
  \label{eq:comment:1}
  p = \frac {M\, g}{a^2} \cdot  z_1.
\end{equation}
Here, $a = 206'=2472''$ is an approximate length of the side of the building.
We get ${_1p} = 0.025$ KSI for WTC 1, and ${_2p} = 0.05$ KSI for WTC 2,
which is three orders of magnitude smaller than nominal 36-100 KSI (ultimate 58-110 KSI)
the vertical columns were able maintain while yielding in plastic deformation.
Bazant \etal\cite{bazant2006} argued that an avalanche propagating through
the primary zone would get sufficiently compacted so that it
could provide necessary pressure.
We see two insurmountable problems with this suggestion.
First, the avalanche front can only ``grow'' thicker - it cannot
expand laterally in such a fashion that would allow its edges to be strong enough to
crush the vertical columns.
Second, for compaction to happen the floor material has to be compressed between two
solid surfaces, and we see that there are no such surfaces on either end of the avalanche
front.
In fact, the strength of the vertical columns will redirect the avalanche (which now
consists only of destroyed floor material) to the region in-between the columns.
The formation of such avalanche is promoted by the relative
weakness of the floors, the resistive force $f$ of which is
$f \sim 0.02$,\cite{nist:Lewetal2003} per each floor,
as compared to the resistive force of the intact
vertical columns, $f_{CC}+f_{PC} \simeq 0.8 + 2.7\cdot z$.
\item[2.] {\bf Rigidity assumption:}
The NIST report claims that the collapse started because the vertical columns
could not absorb the energy of the falling top section of the building.~\cite{nist:2005}
By design, all vertical columns were continuous structures that stretched
from the ground floor to the top of the building.
Lateral support was added to them to prevent them from buckling under
load, so that they would behave as ``short columns.''
For our models we assumed that the vertical columns are indeed short columns:
under compression they maintain their ultimate strength until the yield strain
is reached.
The rigidity assumption enters here as the location where the fracture occurs -
at (according to Bazant\etal\cite{bazant2006}) or near
the interface between the avalanche and the vertical column.
However, this is a slow compression of the column (the velocity of the source
of compression is much smaller than the sound velocity in the steel) so
the stress has time to propagate throughout the whole column causing
the strain to do the same.
As a result, the fractional distance $\lambda_1$ should be
applied to the full length of the column ($\sim H$, the height of the building)
and not to the storey height $\Delta H = H/F_T$.
Actual yield strain $\lambda_y$ can be estimated from the
yield force, $f(z) = r + s \, z$, where a local contribution to the
yield strain, $\delta\lambda_y$, is reversely proportional
to the local yield force, $\delta\lambda_y = C/f(z)$,
in accord with the uniform distribution of deformation energy.
This gives
\begin{equation}
  \label{eq:comment:2}
  \lambda_y = \sum \delta\lambda_y \simeq
  \int_{z_1}^1 \, dz \ \frac{C}{f(z)} =
  \lambda_1 \cdot \frac {r}{s} \cdot \log\frac{r+s}{r+s\,z_1}.
\end{equation}
Here, the limiting case of $s \rightarrow 0$ gives $C = \lambda_1 \, r$,
in which case $\lambda_y = \lambda_1 \cdot (1-z_1)$.
Calculation with intact parameters, $r=0.8$ and $s=2.7$, gives
${_1\lambda}_y = 0.063$ ($z_1 = 1 - F_1/F_T = 0.15$, for WTC 1 ) and
${_2\lambda}_y = 0.046$ ($z_1 = 0.3$, for WTC 2).
We leave as an exercise to the reader to show that these distances are sufficient
to stop the fall of the top section even if one makes a radical assumption that the
avalanche propagated through the primary zone without resistance ($r=s=0$).

Solution:
In the elastic case $(\epsilon\equiv1$) the velocities $v_1$ at the exit of the primary
zone are found from the energy conservation, ${_1v}_1 = 0.42$ and ${_2v}_1=0.5$
(for $\epsilon\equiv0$, one would use $P^2$ as an integral of motion, where $P$
is the momentum of the avalanche).
Let $\lambda\,H$ be the length needed to stop the fall, and let $r=0.8$ and
$s=2.7$ for the intact secondary zone, as before. Then, $\lambda$ is
given by
\begin{equation}
  \label{eq:comment:3}
  \lambda = \frac{z_1 \, v_1^2}{2 \,( f(z_1) - z_1)}.
\end{equation}
We find ${_1\lambda} = 0.013$ for WTC~1, and ${_2\lambda} = 0.029$ for WTC~2,
which are considerably smaller then their yield strains $\lambda_y$'s.
Thus, contrary to the NIST claim, the total plastic deformation of the
intact vertical columns in the secondary zone was more than sufficient to
arrest the fall of the top section.
\end{itemize}
From our discussion so far it follows that $(i)$, the secondary zones of both WTC 1 and
WTC 2 were not intact - in agreement with the hypothetical WMD destroying the
CCs discussed earlier, and $(ii)$, the destruction of the remaining vertical
columns (PCs) was either not through compression, or there
had to be a mechanism present that would pull the PCs inwards and into the path of the
avalanche.

\begin{figure}[h]
  \singlespacing
  \centering
  \includegraphics[scale=0.5,clip]{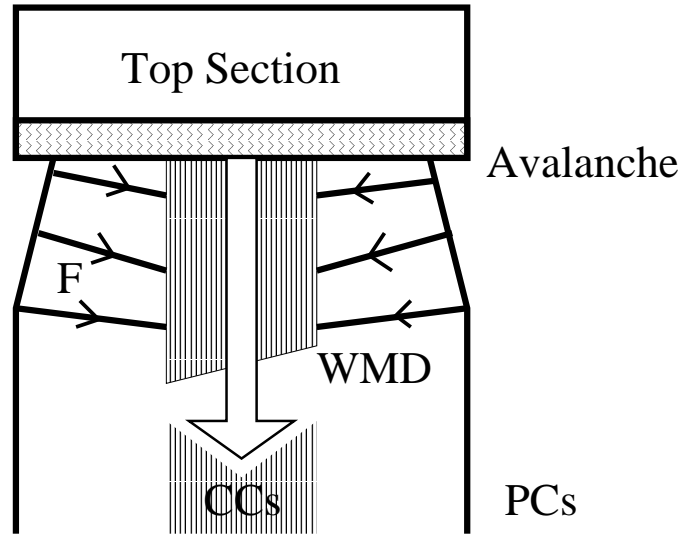}
  \caption{%
    \label{fig:4}
    Hypothetical model of progressive collapse in WTC 1 and 2, caused
    by compromising the central core columns (CCs) at predetermined
    locations along the height of the building and times, labeled
    ``the wave of massive destruction'' (WMD).
    As the core columns (CCs) are disabled, the collapse is opposed by the
    perimeter columns (PCs) mainly,  as discussed in text.
  }
\end{figure}

This said, let us propose a consistent hypothetical model of an avalanche.
The avalanche is created by severing the core columns (CCs) at some distance
from the primary zone.
This makes the avalanche consist of the intact top section, the intact CCs of
which penetrate the secondary zone, and so give it an overall wedge-like appearance.
As a result of weight redistribution, the avalanche now interacts only with the
perimeter columns (PCs) from the top of the primary zone down to the level
at which the CCs were severed from the secondary zone CCs.
The avalanche's CCs pull the secondary zone PCs inwards, and so compromise them,
while the intact top section finishes the PCs as it goes down.
In this way, the avalanche never encounters the rigidity of the whole building,
just of its small section, as discussed earlier.
Furthermore, the pulling action is realized with the intact floor structure in the secondary
zone, through the tension of the floor trusses.
As is known, the tension yield of the floor trusses is much greater then
their shear yield force.
From the outside, it appears as if the avalanche starts at the weakest point of
the remaining structure: the compromised PCs in the primary zone.
By propagating so, the avalanche sees mostly the resistive force of the PCs in the
secondary zone, and some friction from the penetration of avalanche CCs into the
floor structure of the secondary zone.
The compromising of the secondary zone CCs continues so that the next severing point
is always ahead of the avalanche: otherwise, the avalanche's CCs might interfere
with the severing, which if prevented would result in a slowing down
of the avalanche.
The process continues until the avalanche reaches the ground floor.
We show the schematic of such collapse in \fig{fig:4}.

\newpage

\appendix
\section{Collapse of WTC 7}
\begin{figure}[h]
  \singlespacing
  \centering
  \includegraphics[scale=0.4,clip]{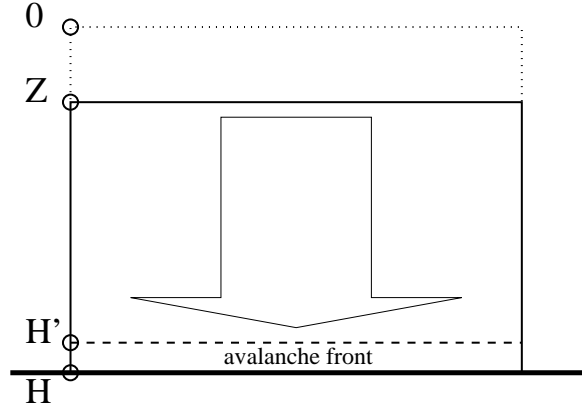}
  \caption{%
    \label{fig:5}
    ``Crush-up'' model of collapse of a building.
    The structural strength of the building fails along its whole size,
    as a result of which the building between $Z=0$ and $Z=H'$
    starts to move in a ``free fall''-like fashion.
    The avalanche front is formed at the height $H'$ and its position with respect to
    the ground level remains fixed for the whole duration of collapse.
  }
\end{figure}

World Trade Center 7 was a building of height $H=186$ m and had total
number of $F_T=47$ stories.
It perished together with WTC 1 and 2,
where its collapse lasted at most ${_7T}_c = 6.5$ s.

\subsection{Mathematical Models}

We use Bazant \etal's\cite{bazant2006} term ``crush-up'' to describe
the collapse sequence shown in \fig{fig:5}.
Assuming closed system, all the relevant energies can be described in terms
of position of the top of the building, $Z$.
The mass of the moving part is $m(Z) = \int_Z^{H'}\, dX \ \mu(H'-X)$,
where $\mu=\mu(Z)$ is the mass distribution in the building.
For brevity, we assume that $\mu(Z) = \mu = M/H$ is a constant, where $H$ is the
height of the building and $M$ is its total mass.
The kinetic energy is then $K(Z,\dot Z) = \frac 1 2 \, \mu (H'-Z) \dot Z^2$,
while the potential energy is
$U(Z)=-\int_Z^{H'}\ dX\ \mu \, g\, X - \int_0^Z\ dX\ \mu \, g \, H'
= \frac 1 2 \, \mu\,g\,(Z-H')^2 - \mu\,g\,H'^2$,
where the first term comes from the part of the building still standing,
while the second term comes from the collapsed part of the building
at rest at height $H'$.
On the other hand, the structural energy is $L = \int_Z^{H'}\ dX\ f(X-Z) =
M\,g\,\left(r \,(H'-Z) + \frac s {2\,H} (H'-Z)^2 \right)$.
As was done before, the equations of motion are derived from the Lagrangian
${\cal L} = T-U-L$.
The equation of motion for the dimensionless position $z = Z/H$ in dimensionless
time $\tau = t/T$, where $T=\sqrt{2\,H/g}=6.16$ s is the free-fall time,
is given by
\begin{equation}
  \label{eq:cd:all}
  z''(\tau) = 2\,(1-s^*) - \frac{2\,r^*}{\delta - z(\tau)}
  + \left(1 - \frac \epsilon 2\right)\, \frac {z(\tau)'^2}{\delta -z(\tau)}
  \quad,
\end{equation}
where $\delta = H'/H \le 1$ is the position of the avalanche front.
The parameters of the building in the continuous model, $(r^*,s^*)$, are related to
the discrete (nominal) values of the building $(r,s)$, as before,
$r^* \simeq 0.25 \, r$ and $s^* \simeq 0.25\, s$, cf. \eqns{eq:rc:all} and \noeqn{eq:rs:wtc}.
Assuming $\epsilon\equiv0$ in \eqn{eq:cd:all} yields Bazant \etal's
model\cite{bazant2006} without their boost factor.

We observe that $(i)$, in \eqn{eq:cd:all} the parameter $\delta$ depends
on the external circumstances
of the collapse, and $(ii)$, as $z\rightarrow\delta$ the acceleration diverges.
This divergence comes from the fact that as the building gets shorter,
it maintains non-zero resistive force.
As this is not the case in an actual building, we fix that problem by
recognizing that the acceleration of vanishing building cannot be
greater than $g\,(1 - s^*)$.
The equation of motion we are thus solving reads
\begin{equation}
  \label{eq:cd:full}
  z''(\tau) =
  \left\{
    \begin{array}{ll}
      2\,(1-s^*) - \frac{2\,r^*}{\delta - z(\tau)}
      + \left(1 - \frac \epsilon 2\right)\, \frac {z(\tau)'^2}{\delta -z(\tau)}
      \quad, &\mbox{if this is less than}\quad 2\,(1-s^*),\\
      2 \, (1 - s^*), &\mbox{otherwise.}
    \end{array}
  \right.
\end{equation}
As we have shown for WTC 1 and WTC 2, mathematical models of progressive collapse
are fairly robust regarding their discrete or continuous and plastic or elastic nature.
For the brevity of presentation, we base all our results on the analysis of continuous models.
We use the spread between plastic and elastic model to find the range of $(r,s)$ that
yields the observed collapse time.

\subsection{Failure of the NIST/FEMA scenario}

We formulate the NIST/FEMA scenario regarding the collapse of WTC 7
as follows:\cite{fema:2002,nist:2005}
The conditions in the building gradually worsened throughout
the whole height because of the heat from unattended fires.
Eventually the building failed, where the avalanche front was formed
between the floors $F_0=5$ and $F_1=7$.

As the actual structural parameters of WTC 7 are not available, to the best of
our knowledge, we base the following analysis on known properties a building of
such proportions typically has, namely, its total factor of safety:
for intact building we take this to be 3.5, just like WTC 1 and 2.
Thus, $r$ and $s$ in the intact building are somewhere on the line $r+s \sim 3.5$.
We observe that due to prolonged heat from unattended ambient fires
the structural strength of steel can be reduced by at most factor
of $\nu\sim2$, corresponding to 600$^o$C.
However, as the heat damage is a function of maximal temperature of the fires rather
than their duration, we take for $\nu$ more conservative value of $\nu\sim1.5$.
Thus, $r$ and $s$ in WTC 7 following the heat damage are above the line
$r+s \sim  2.3$ (=3.5/1.5).

\begin{figure}[htp]
  \singlespacing
  \centering
  \includegraphics[scale=0.7,clip]{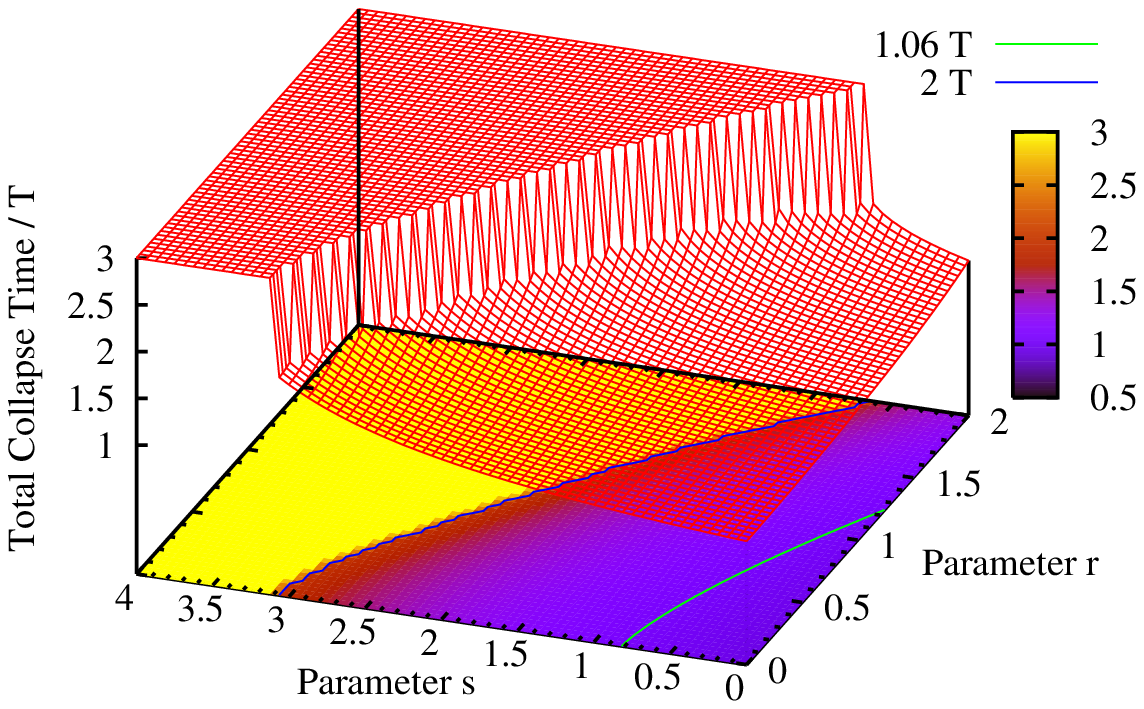}
  \includegraphics[scale=0.7,clip]{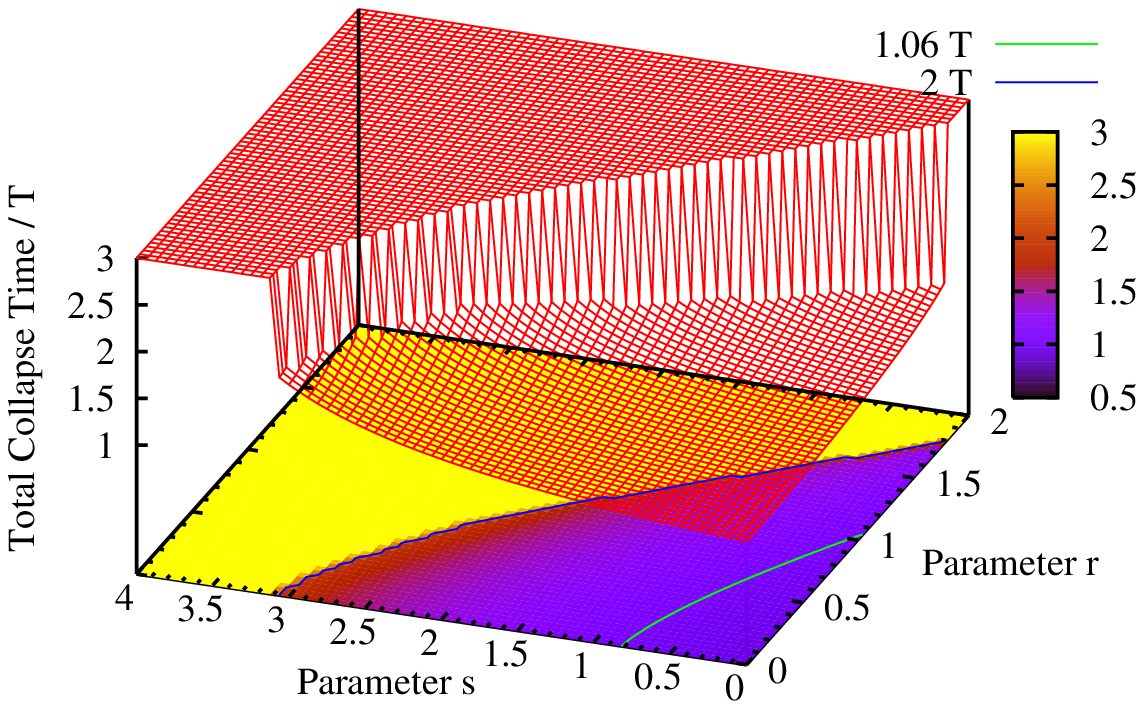}
  \caption{%
    \label{fig:6}
    Duration of collapse $\tau_c=\tau_c(r,s)$ in plastic (left) and elastic (right)
    model, given in \eqn{eq:cd:full}.
    The base of the plot contains two contours, $\tau_c(r,s)=2$ and $\tau_c(r,s)=1.06$, where the
    latter corresponds to the observed collapse time of $T_c=6.5$ s.
  }
\end{figure}

Consider now that the avalanche front in WTC 7 formed at the
floor $F_0$, at the height $H'$, where $1-H'/H = \delta = 1 - F_0/F_T = 0.9$.
This allows us to construct the (static) collapse initiation line,
\begin{equation}
  \label{eq:cd:cil}
  \frac {r}{\delta} + s = 1.
\end{equation}
The parameters $(r,s)$ in the damaged building have to be below the collapse initiation
line, \eqn{eq:cd:cil}, for building to collapse.
We check the consistency of static and dynamic features of collapse as before,
by computing the collapse duration times from elastic ($\epsilon\equiv1$)
and plastic ($\epsilon\equiv0$) mathematical model~\noeqn{eq:cd:full},
where we use standard initial conditions, $z(0)=z'(0)=0$.
We show the collapse duration as a function of $r$ and $s$
in~\fig{fig:6}, where we also plot two contours,
$\tau_c(r,s) = 2$ and $\tau_c(r,s) = 1.06$.

\begin{figure}[htp]
  \singlespacing
  \centering
  \includegraphics[scale=0.4,clip]{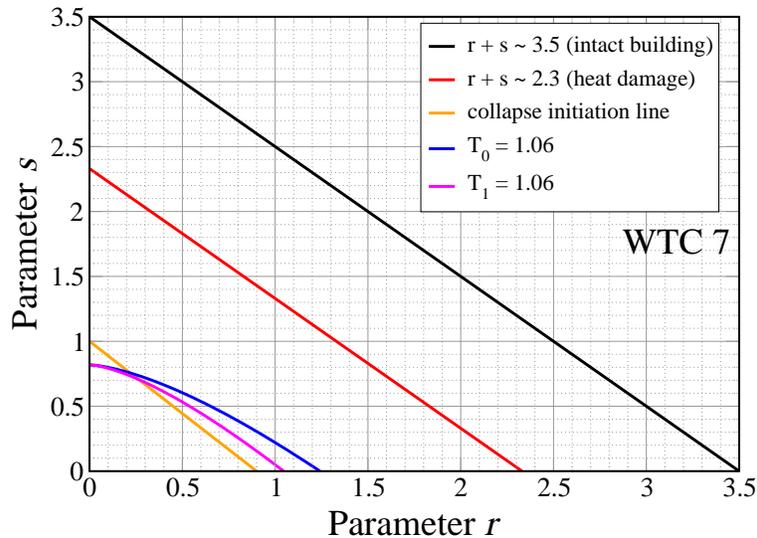}
  \caption{%
    \label{fig:7}
    As a result of gradually worsening conditions in WTC 7, its $r$ and $s$
    change from being intact, $r+s \sim 3.5$ (black), to being compromised
    by heat, $r+s \sim 2.3$ (red).
    The collapse initiation line (orange) is determined from the height at which the
    avalanche front forms during the ``crush-up,'' and it agrees
    with the computed collapse time
    ($T_1$ in elastic model, pink; and $T_0$ in plastic model, blue).
    Thus, for WTC 7 to collapse yet another damaging event is necessary,
    because of which its $r$ and $s$ decrease from being close to the heat damage
    line (red) to being close to the overlap of the collapse initiation (orange)
    and the duration (blue,pink) line.
  }
\end{figure}

Our findings are summarized in~\fig{fig:7} where we show possible
$r$ and $s$ of WTC 7 together with its collapse initiation and the duration line.
There can be seen that the collapse initiation line (orange)
is well below the line that describes the building damaged by heat (red).
We note that both of these lines are derived from the static properties
of the building.
On the other hand, the collapse duration line (pink and blue, for elastic
and plastic model, respectively) overlaps with the collapse initiation line (orange),
meaning that the two are mutually consistent.
If the NIST/FEMA scenario were accurate description of transition to collapse
then the collapse initiation line had to coincide with the heat damage line.
Instead, judging by the gap between the two lines, there had to exist
yet another damaging event as a result of which
the building  collapsed.


\newpage
% references
\bibliography{%
  bibliography/applied%
}

\end{document}